\documentclass[12pt]{article}
\usepackage{amssymb}
\usepackage{graphicx}

\makeatletter
 \@addtoreset{equation}{section}
 
\makeatother

\begin{document}


\begin{titlepage}

\renewcommand{\thefootnote}{\fnsymbol{footnote}}

\begin{flushright}
\begin{tabular}{l}
UTHEP-583\\
RIKEN-TH-155
\end{tabular}
\end{flushright}

\bigskip

\begin{center}
{\Large \bf Light-Cone Gauge 
Superstring Field Theory and Dimensional Regularization}
\end{center}

\bigskip

\begin{center}
{\large
Yutaka Baba}${}^{a}$\footnote{e-mail:
        ybaba@riken.jp},
{\large Nobuyuki Ishibashi}${}^{b}$\footnote{e-mail:
        ishibash@het.ph.tsukuba.ac.jp},
{\large Koichi Murakami}${}^{a}$\footnote{e-mail:
        murakami@riken.jp}
\end{center}

\begin{center}
${}^{a}${\it Theoretical Physics Laboratory, RIKEN,\\
         Wako, Saitama 351-0198, Japan}
\end{center}
\begin{center}
${}^{b}${\it
Institute of Physics, University of Tsukuba,\\
Tsukuba, Ibaraki 305-8571, Japan}\\
\end{center}

\bigskip

\bigskip

\bigskip

\begin{abstract}
We propose a dimensional regularization scheme in the light-cone
gauge NSR superstring field theory to regularize the divergences
caused by the colliding supercurrents inserted at the interaction
points. We study the tree amplitudes and show that
our scheme actually regularize the divergences.
We examine the four-point amplitudes for the NS-NS closed
strings and find that the results in the first-quantized theory
are reproduced without any counterterms.

\end{abstract}

\setcounter{footnote}{0}
\renewcommand{\thefootnote}{\arabic{footnote}}

\end{titlepage}

\section{Introduction}

In the light-cone gauge closed NSR superstring field theory
there exists the joining-splitting
type cubic interaction.
In order that the theory should be Lorentz invariant,
the transverse supercurrent must  be inserted at
the interaction point \cite{Mandelstam:1974hk}\cite{Sin:1988yf}.
In calculating the amplitudes perturbatively,
these supercurrents give rise to unwanted divergences
when they get close to each other.
Since the Green-Schwarz formulation is equivalent to 
the NSR formulation in the light-cone gauge \cite{Mandelstam:1985wh}, 
the same difficulty exists in the light-cone gauge
string field theory of Green-Schwarz 
strings \cite{Greensite:1986gv}\cite{Green:1987qu}.\footnote{
  Similar problems in covariant superstring field theories 
  are studied in \cite{Wendt:1987zh}.}

In order to make these theories well-defined even classically,
we have to regularize these divergences and 
add counterterms to cancel them. 
The results from the supersheet 
approach~\cite{Berkovits:1985ji}\cite{Aoki:1990yn} 
imply that the divergences are formally from 
the integration of total derivative terms over the moduli space. 
Therefore, by choosing a good regularization, 
we may be able to tame the divergences. 
Usually in the literature, 
point splitting type regularizations are employed.
In the field theory of  point particles,
the dimensional regularization is very powerful.
In this paper, we therefore pursue this regularization 
in the light-cone gauge NSR superstring field theory to
regularize  the  divergences of the amplitudes
originating in the colliding supercurrents. 

In string theory, 
dimensional regularization can be implemented by 
shifting the number of space-time dimensions, 
or shifting the central charge of the conformal field theory 
on the worldsheet. 
In this paper, 
we consider the worldsheet CFT which consists of the 
ordinary transverse coordinate fields and 
a superconformal field theory with Virasoro central charge 
$c^{\mathrm{ext}}$. 
It is possible to construct light-cone gauge string field theory 
with such worldsheet CFT, although it is not Lorentz invariant. 
We study the tree amplitudes and show that this procedure
actually regularizes the above mentioned divergences.
The question is  what kind of counterterms are needed
to cancel the divergences to obtain the amplitudes which 
coincide with the ones derived by using the first-quantized 
formalism.
In this paper, we examine the four-point tree amplitudes
for the NS-NS closed strings and show that no counterterms
are needed in this case.

The organization of this paper is as follows.
In the next section, we study the $N$-string tree
amplitudes of the light-cone gauge closed string field theory
for NSR superstrings with extra CFT,
and see that the tree amplitudes become well-defined
by making the central charge $c^{\mathrm{ext}}$ of the CFT 
on the worldsheet largely negative.
Therefore we can define the amplitudes for such $c^{\mathrm{ext}}$ and 
analytically continue $c^{\mathrm{ext}}$ to $c^{\mathrm{ext}}=0$.
In section \ref{sec:covariant}, 
in order to compare the results with those of the first-quantized 
formalism, we rewrite the NS-NS tree amplitudes 
by introducing the longitudinal coordinates and the ghosts. 
For $c^{\mathrm{ext}}=0$, this was done in Ref.~\cite{Aoki:1990yn}
to show the equivalence between the light-cone gauge amplitudes 
and the covariant ones. 
We essentially follow their procedure in the component formalism.
In section \ref{sec:4pt},
we show that the four-point tree amplitudes for
the NS-NS closed strings are finite in our regularization scheme 
and they coincide with the results in the first-quantized formalism. 
Therefore no four-string contact interaction terms are needed 
as counterterms.
Section \ref{sec:discussion} is devoted to discussions.
In appendix \ref{sec:action}, we summarize our convention for 
the light-cone gauge string field theory.
In appendix \ref{sec:Mandelstam}, the Mandelstam mapping is given.
In appendix \ref{sec:exp-Gamma},
a derivation of the oscillator independent
part of the tree interaction vertex for $N$ strings
is presented.

\section{Light-cone Gauge Superstring Amplitude}
\label{sec:lc-amp}

Let us consider the light-cone gauge closed string field theory 
of NSR superstrings whose worldsheet theory consists of 
the usual CFT for the NSR superstring
in the ten dimensional space-time 
and a superconformal field theory 
with central charge $c^{\mathrm{ext}}$.
We will refer to the latter as the extra CFT or extra sector.
Even with $c^{\mathrm{ext}}\neq 0$, one can define the light-cone gauge
string field theory using the super Virasoro operators, 
although the ten dimensional Lorentz invariance is broken. 
We calculate the amplitudes perturbatively,
and the extra sector part in every external state
of the amplitudes is always taken to be the conformal vacuum.  
In this paper, we restrict ourselves to the tree amplitudes. 
Since the relevant property of the extra CFT is only its
central charge in this situation, we do not specify the details
of the extra CFT.
While we concentrate on the case that  
all the external strings
belong to the NS-NS sector,
it is not difficult to extend our calculation into the case
where the Ramond  strings are involved.

The $N$-point NS-NS tree amplitudes 
in the light-cone gauge string field theory
are calculated from the action 
defined in appendix \ref{sec:action}.
They are given as integrals over the moduli space,
\begin{equation}
\mathcal{A}
 = (ig)^{N-2} C_N 
 \int \left( \prod_{I} \frac{ d^{2} \mathcal{T}_{I} }{4\pi} \right)
 F (\mathcal{T}_{I},\bar{\mathcal{T}}_{I})~,
\label{eq:amplitude}
\end{equation}
where $g$ denotes the string coupling,
$C_N$ is the symmetric factor,
$\mathcal{T}_{I}$ are the $N-3$ complex moduli parameters
of the string diagram defined as
\begin{equation}
\mathcal{T}_{I} = \rho (z_{I+1}) - \rho (z_{I})
= T_I + i \alpha_m \theta_I ~,
 \qquad 
\alpha_m>0~,
\end{equation}
and
\begin{equation}
 \int \frac{d^2 \mathcal{T}_I}{4\pi}
=-i \alpha_m \int_0^{\infty} dT_I
\int_{0}^{2\pi} \frac{d\theta_I }{2\pi}~.
\end{equation}
Here $\rho (z)$ is the Mandelstam mapping \cite{Mandelstam:1973jk}
given in eq.(\ref{eq:Mandelstam}),
 $z_I$ denote the coordinates of the $N-2$ interaction points
on the complex $z$-plane defined in appendix \ref{sec:Mandelstam}
and $\alpha_m$ denote the string-length parameters
of the intermediate strings. 
The propagator for the intermediate strings  
can be written as 
\begin{equation}
-i 
\int \frac{d^2 \mathcal{T}_I}{4\pi}
e^{- \frac{\mathcal{T}_I}{\alpha_m}
   \left( L_0^{\mathrm{LC} }-\frac{c}{24} \right)
   - \frac{\bar{\mathcal{T}}_I}{\alpha_m}
\left( \tilde{L}_0^{\mathrm{LC}} - \frac{c}{24}\right) }
      |R(m,m')\rangle
 \frac{1}{-i\alpha_m}
 ~,
\qquad \mbox{with }\alpha_m>0~,
\label{eq:propagator}
\end{equation}  
where $c$ is the total central charge of this system,
\begin{equation}
c=c^{(X,\psi)\mathrm{LC}} + c^{\mathrm{ext}}~,
\qquad
c^{(X,\psi) \mathrm{LC}}=12~.
\label{eq:center}
\end{equation} 
Connecting the three-string vertices using 
the propagator (\ref{eq:propagator}), 
we obtain the $N$-string interaction vertex
$\langle V_{N}|$, which takes the form
\begin{equation}
\langle V_{N}| = 
4\pi \delta \left( 
  \sum_{r=1}^N \alpha_r
\right)
\langle V_{N}^{\mathrm{LPP}}|
   e^{-\Gamma^{\mathrm{LC}}}
    \prod_{I} \left( \partial^{2} \rho (z_{I})
                  \bar{\partial}^{2} \bar{\rho}(\bar{z}_{I})
           \right)^{-\frac{3}{4}}
          T^{\mathrm{LC}}_{F} (z_{I})
          \tilde{T}_{F}^{\mathrm{LC}} (\bar{z}_{I})~.
\label{eq:vertex-N}
\end{equation}
Here $\langle V_{N}^{\mathrm{LPP}}|$ is the LPP vertex
that is determined by the correlation functions of the
worldsheet theory through the prescription of
LeClair, Peskin and Preitschopf~\cite{LeClair:1988sp}.
We take the normalization of this vertex as
\begin{equation}
\int \left( \prod_{r=1}^N
 \frac{d^8p_r^i}{(2\pi)^8}\right)
\langle V_{N}^{\mathrm{LPP}}|0 \rangle_1
 \left( \prod_{r=2}^N
 |0\rangle_r (2\pi)^8 \delta^8(p_r^i)
 \right)=1~.
\end{equation} 
$T^{\mathrm{LC}}_{F}$ denotes the transverse supercurrent of
the worldsheet theory given in eq.(\ref{eq:Tf}).
$e^{-\Gamma^{\mathrm{LC}}}$ is the value of
the $N$-point amplitude  without any supercurrent insertions 
when all the external states 
are taken to be the conformal vacua.

The overlap of the LPP vertex with external states can 
be related to the correlation functions on the complex $z$-plane 
via the Mandelstam mapping $\rho (z)$.
Therefore, the integrand $F(\mathcal{T}_{I}, \bar{\mathcal{T}}_{I})$
of the amplitude (\ref{eq:amplitude}) can be expressed by
the path integral,\footnote{
  We are not interested in overall numerical factors
  of the amplitudes. The symbol $\sim$ will be used to indicate
  the equality up to a constant factor.}
\begin{eqnarray}
F (\mathcal{T}_{I}, \bar{\mathcal{T}}_{I})
 &\sim&         (4\pi)
 \delta \left( \sum_{r=1}^N \alpha_r
           \right)(2\pi)
  \delta \left( \sum_{r=1}^N p_r^-
           \right)
\nonumber\\
&&\times
 \int 
 [dX^{i} d\psi^{i} d\phi^{\mathrm{ext}}] \,
 e^{-S^{\mathrm{LC}}-\Gamma^{\mathrm{LC}}}
 \prod_{r=1}^{N} \left[
            \alpha_r V_{r}^{\mathrm{LC}} (w_{r}=0,\bar{w}_{r}=0)
            e^{-p^{-}_{r} \tau_{0}^{(r)}}
          \right]
\nonumber\\
 &&   \times
 \prod_{I} \left[ \left( \partial^{2} \rho (z_{I})
                  \bar{\partial}^{2} \bar{\rho}(\bar{z}_{I})
           \right)^{-\frac{3}{4}}
          T^{\mathrm{LC}}_{F} (z_{I})
          \tilde{T}_{F}^{\mathrm{LC}} (\bar{z}_{I})
           \right]~.
\label{eq:F-LC}
\end{eqnarray}
Here $\tau_{0}^{(r)}$ is defined
in eq.(\ref{eq:rho-zIr}),
and $w_{r}$ is the coordinate of the unit disk of the $r$-th
external string given in appendix \ref{sec:Mandelstam},
the origin of which corresponds to the puncture at $z=Z_{r}$
on the $z$-plane.
$\phi^\mathrm{ext}$ denotes the fields in the extra sector
and $S^{\mathrm{LC}}$ is the action of
the worldsheet theory of the light-cone gauge NSR superstring
containing the extra sector.
$V^{\mathrm{LC}}_{r} (w_{r},\bar{w}_r)$ denotes the vertex operator
of the $r$-th string corresponding to the string state
\begin{equation}
\epsilon^{(r)}_{\{ij\}}
\alpha^{i_{1} (r)}_{-n_{1}} \cdots
\tilde{\alpha}^{\tilde{\imath}_{1} (r)}_{-\tilde{n}_{1}} \cdots
\psi^{j_{1} (r)}_{-s_{1}} \cdots
\tilde{\psi}^{\tilde{\jmath}_{1} (r)}_{-\tilde{s}_{1}} \cdots
 |p^{i}_{r} \rangle_r~,
\label{eq:stringstate-LC}
\end{equation}
namely
\begin{eqnarray}
V^{\mathrm{LC}}_{r} (w_{r},\bar{w}_{r})
&=&
\epsilon^{(r)}_{\{ij\}}
\oint_{w_r} \frac{dw_{n_1}}{2\pi i}
\frac{i \partial X^{i_{1}(r) }(w_{n_1})}
      {(w_{n_1})^{n_1}} 
 \cdots
 \oint_{\bar{w}_r} \frac{d\bar{w}_{\tilde{n}_1}}{2\pi i}
\frac{i \bar{\partial} X^{\tilde{\imath}_{1}(r) }
          (\bar{w}_{\tilde{n}_1})}
     {(\bar{w}_{\tilde{n}_1})^{\tilde{n}_1}} 
 \cdots
 \nonumber\\
&& \times
 \oint_{w_r} \frac{dw_{s_1}}{2\pi i}
\frac{ \psi^{j_{1} (r)} (w_{s_1})}
      {(w_{s_1})^{s_{1} + \frac{1}{2}}}
 \cdots
  \oint_{\bar{w}_r} \frac{d\bar{w}_{\tilde{s}_1}}{2\pi i}
\frac{ \tilde{\psi}^{\tilde{\jmath}_{1} (r)} (\bar{w}_{\tilde{s}_1})}
      {(\bar{w}_{\tilde{s}_1})^{\tilde{s}_{1} + \frac{1}{2}}}
 \cdots
 e^{i p^{i}_{r} X^{i(r)}} (w_{r},\bar{w}_{r})~,
\label{eq:VrLC}
\end{eqnarray}
where $\epsilon^{(r)}_{\{ij\}}$ is a shorthand notation of
the polarization tensor
$\epsilon^{(r)}_{i_{1}\cdots j_{1} \cdots \tilde{\imath}_{1}
                 \cdots \tilde{\jmath}_{1} \cdots}
  ( p^{i}_{r})$.
For later use, we introduce the mode numbers $N_{r}$ and
$\tilde{N}_{r}$ of this vertex operator defined as
\begin{equation}
N_{r} = \sum_{l} n_{l} + \sum_{l}s_{l}~,
\qquad
\tilde{N}_{r} = \sum_{l} \tilde{n}_{l} + \sum_{l} \tilde{s}_{l}~.
\label{eq:levelnumber}
\end{equation}
The level-matching condition and 
the on-shell condition\footnote{
  One may consider that 
  the on-shell condition should be
  $p_r^\mu p_{\mu r} + N_r+ \tilde{N}_r -\frac{c}{12}=0$
  rather than eq.(\ref{eq:onshell}).
  The difference between these two on-shell conditions has
  no effect in the limit $c^{\mathrm{ext}} \to 0$. 
}
require that
\begin{equation}
N_{r} = \tilde{N}_{r}~,
\qquad
p_r^\mu p_{\mu r} + N_r + \tilde{N}_r -1=0~.
\label{eq:onshell}
\end{equation}
We note that
the factor $\alpha_r$ in front of each vertex operator
$V_r^{\mathrm{LC}}$ in the integrand
in eq.(\ref{eq:F-LC}) comes from that contained in the measure
$dr$ defined in eq.(\ref{eq:measure}).

$e^{-\Gamma^{\mathrm{LC}}}$ can be obtained by evaluating
the Liouville action for the flat metric on the light-cone
diagram \cite{Mandelstam:1985ww} as briefly illustrated in
appendix \ref{sec:exp-Gamma}, and we have
\begin{equation}
e^{-\Gamma^{\mathrm{LC}}}
=  
\mathrm{sgn}
\left(
\prod_{r=1}^{N} \alpha_{r}
     \right)
\left| \prod_{r=1}^{N} \alpha_{r}
     \right|^{-\frac{c}{12}}
     \left| \sum_{s=1}^{N} \alpha_{s} Z_{s}
     \right|^{\frac{c}{6}}
     e^{-\frac{c}{12} \sum_{r=1}^{N}
      \mathrm{Re} \bar{N}^{rr}_{00} 
     }
    \prod_{I} \left| \partial^{2} \rho (z_{I})
             \right|^{-\frac{c}{24}}~.
\label{eq:exp-Gamma}
\end{equation}
Here 
$\bar{N}^{rr}_{00}$ is a Neumann coefficient
given by
\begin{equation}
\bar{N}^{rr}_{00}
 = - \sum_{s\neq r} \frac{\alpha_{s}}{\alpha_{r}}
   \ln (Z_{r} - Z_{s}) + \frac{\tau^{(r)}_{0}+i\beta_{r}}
                              {\alpha_{r}}~,
\end{equation}
and $\beta_{r}$ is defined in eq.(\ref{eq:rho-wr}).

Combining eqs.(\ref{eq:F-LC}) and (\ref{eq:exp-Gamma}),
we obtain
\begin{eqnarray}
F(\mathcal{T}_{I},\bar{\mathcal{T}}_{I})
&\sim& 
(4\pi) \delta
     \left( \sum_{r=1}^{N} \alpha_{r}
    \right)
(2\pi) \delta
     \left( \sum_{r=1}^{N} p_{r}^-
    \right)
\int [dX^{i}d\psi^{i} d \phi^{\mathrm{ext}}] 
  e^{-S^{\mathrm{LC}}}
\left| \mathcal{D} \right|^{2c^{\mathrm{ext}}}
    \left| \sum_{s=1}^{N} \alpha_{s} Z_{s}
    \right|^{2}
\nonumber\\
&& \times
     \prod_{r=1}^{N} \left[
        V^{\mathrm{LC}}_{r} (w_{r}=0,\bar{w}_{r}=0)
        e^{-p_{r}^{-} \tau^{(r)}_{0}}
        e^{- 
            \mathrm{Re}\bar{N}^{rr}_{00} }
        \right]
\nonumber\\
&& \times \prod_{I} \left[
    \left( \partial^{2} \rho (z_{I})
           \bar{\partial}^{2} \bar{\rho} (\bar{z}_{I})
    \right)^{-1 }
    T_{F}^{\mathrm{LC}} (z_{I})
    \tilde{T}^{\mathrm{LC}}_{F} (\bar{z}_{I}) \right]~,
\label{eq:F-LC2}
\end{eqnarray}
where $\mathcal{D}$ is defined as
\begin{equation}
\mathcal{D} = 
\left|\prod_{r=1}^N \alpha_r \right|^{-\frac{1}{24}}
 \left( \sum_{s=1}^{N} \alpha_{s} Z_{s}
    \right)^{ \frac{1}{12}}
     e^{- \frac{1}{24} \sum_{r=1}^{N}
             \bar{N}^{rr}_{00}}
    \left( \prod_{I}
    \partial^{2} \rho (z_{I})
    \right)^{-\frac{1}{48}}~.
\label{eq:D}
\end{equation}

In general, $F(\mathcal{T}_{I},\bar{\mathcal{T}}_{I})$ is 
singular in the limit $z_{I} - z_{J} \rightarrow 0$. 
Taking eq.(\ref{eq:deldelrho}) into account,
we find that if $c^{\mathrm{ext}}$
is taken to be a sufficiently large negative value
as a regularization, the amplitude (\ref{eq:amplitude})
becomes non-singular as $z_{I} - z_{J} \rightarrow 0$.
Other singularities can be dealt with by the analytic continuation 
of the external momenta $p_r$. 
Therefore one can define the integral in eq.(\ref{eq:amplitude}) 
for such $c^{\mathrm{ext}}$ and analytically continue 
$c^{\mathrm{ext}}$ to $c^{\mathrm{ext}}=0$. 
Thus one obtains the dimensionally regularized amplitudes.

While we have studied only the NS-NS closed strings,
it is easy to find that the regularization scheme mentioned above 
works for the amplitudes involving the Ramond strings as well.

\section{Relation to Covariant Formulation}
\label{sec:covariant}

We would like to compare the NS-NS tree
 amplitudes defined in the last section
with  those calculated using the first-quantized formalism.  
In order to do so, it is convenient to recast the
integrand $F(\mathcal{T}_{I},\bar{\mathcal{T}}_{I})$
in eq.(\ref{eq:F-LC2}) into a ``covariant" form, 
by introducing the longitudinal coordinates and the ghosts. 
For $c^{\mathrm{ext}}= 0$, 
such a procedure was given in Ref.~\cite{Aoki:1990yn} using 
the superspace  formalism. 
Here we would like to follow their procedure in the component language.
Since we are dealing with $c^{\mathrm{ext}}\neq 0$ case, 
we cannot obtain Lorentz covariant amplitudes by doing so. 
However, we can obtain the amplitudes which can be compared with the 
covariant ones in the limit $c^{\mathrm{ext}}\to 0$.

\subsubsection*{Ghosts}

Let us first introduce the ghost fields $b,c,\beta ,\gamma$. 
We bosonize the $\beta \gamma$-ghosts in the usual
way~\cite{Friedan:1985ge} as
\begin{equation}
\beta (z) = e^{-\phi } \partial \xi (z)~,
\quad
\gamma (z) = \eta e^{\phi}(z)~.
\end{equation}
On the complex plane in which we are interested here, we 
have the following identity,
\begin{eqnarray}
\lefteqn{
\int [
           db dc 
           d\beta d\gamma ]
      \, e^{-S_{\mathrm{gh}} }
      \left( \lim_{z,\bar{z} \rightarrow \infty}
    \frac{1}{|z|^{4}} c(z) \tilde{c} (\bar{z})
    \right)
} \nonumber\\
&& \quad   \times
\prod_{I} \left[
   b(z_I) \tilde{b}(\bar{z}_I)
   e^{\phi} (z_I) e^{\tilde{\phi} } (\bar{z}_I)\right]
\prod_{r=1}^{N} \left[
   c(Z_r) \tilde{c}(\bar{Z}_r)
   e^{-\phi }(Z_r) e^{-\tilde{\phi} } (\bar{Z}_r)
  \right]
\nonumber\\
&\sim&
\frac{\prod_{r<s}|Z_r-Z_s|^2\prod_{I<J}|z_I - z_J|^2}
   {\prod_{I,r}|z_I - Z_r|^2}
 \times \frac{\prod_{I,r}|z_I - Z_r|^2}
   {\prod_{r<s}|Z_r-Z_s|^2\prod_{I<J}|z_I - z_J|^2}
\nonumber\\
&=& 1~,
\label{eq:ghostsector}
\end{eqnarray}
where $S_{\mathrm{gh}}$ denotes the worldsheet action for
$(b,c;\beta,\gamma)$.
The two factors in the third line in this equation
represent the contributions
from the $(b,c)$ and the $(\beta,\gamma)$
sectors, respectively.
One can notice that the ghost number of the operators
inserted in the path integral (\ref{eq:ghostsector})
is the one that makes this correlation function nonvanishing.

Since the left hand side of eq.(\ref{eq:ghostsector}) is 
just a constant, we can introduce the ghost sector by multiplying 
the right hand side of eq.(\ref{eq:F-LC2}) by this path integral, 
without changing
$F(\mathcal{T}_{I},\bar{\mathcal{T}}_{I})$.

\subsubsection*{Longitudinal coordinates}

Next, let us consider the longitudinal coordinates
$X^{\pm},\psi^{\pm}$.
For the path integral of these fields,
we have
\begin{eqnarray}
\lefteqn{
\int [dX^{\pm}d\psi^{\pm} ]e^{-S_{\pm}}
\prod_{r=1}^{N} 
  V_{r}^{\mathrm{DDF}} (Z_r,\bar{Z}_r) 
}\nonumber\\
&\sim&
 (2\pi)^2 \delta\left(\sum_{r=1}^N p_r^+ \right)
  \delta\left(\sum_{r=1}^N p_r^- \right)
 \prod_{r=1}^{N} 
 \left[
  V_{r}^{\mathrm{LC}} (w_r=0,\bar{w}_r=0) 
e^{- p_r^- \tau_0^{(r)}  - 
\mathrm{Re} \bar{N}_{00}^{rr} }
\right]~.
\label{eq:addsector}
\end{eqnarray}
Here $S_{\pm}$ denotes the worldsheet action
for $(X^{\pm},\psi^{\pm})$. 
$V_r^{\mathrm{DDF}}$ is 
the vertex operator for the DDF state
corresponding to $V_{r}^{\mathrm{LC}}$,
the explicit form of which will be given below.
We can introduce the longitudinal coordinates by substituting 
eq.(\ref{eq:addsector}) into the right hand side of eq.(\ref{eq:F-LC2}). 

In the following, we will prove eq.(\ref{eq:addsector})
by performing the path integral on the left hand side. 
Before doing this,
let us illustrate what $V_r^{\mathrm{DDF}}(Z_r,\bar{Z}_r)$ is. 
The DDF state corresponding to the light-cone gauge
string state (\ref{eq:stringstate-LC}) is given by
\begin{equation}
\epsilon^{(r)}_{\{ij\}}
A_{-n_{1}}^{i_{1} (r)} \cdots
\tilde{A}_{-\tilde{n}_{1}}^{\tilde{\imath}_{1} (r)} \cdots
B_{-s_{1}}^{j_{1} (r)} \cdots
\tilde{B}_{-\tilde{s}_{1}}^{\tilde{\jmath}_{1}(r)}
\cdots
\left| 
  p_{r} + \frac{1}{2} \left( N_{r}+\tilde{N}_{r} \right) k_{r}
\right\rangle_{r}~,
\label{eq:DDFstate}
\end{equation}
where $N_{r}$, $\tilde{N}_{r}$ are the mode numbers given in
eq.(\ref{eq:levelnumber}), $k_{r}$ is the $10$-momentum with
\begin{equation}
k^{-}_{r} = - \frac{1}{p^{+}_{r}}~,
\quad
k^{+}_{r} = k^{i}_{r} =0~,
\end{equation}
the operators $A_{-n}^{i (r)}$ and $B_{-s}^{i (r)}$
are defined as
\begin{eqnarray}
A_{-n}^{i(r)} &=& \oint_{Z_{r}} \frac{dz_n^{(r)}}{2\pi i}
 \left(i \partial X^{i}  + \frac{n}{p_r^+} \psi^i \psi^+\right)
 e^{-i\frac{n}{p^{+}_{r}}
     X^{+}_{L} } (z_n^{(r)})~,
\nonumber\\
B_{-s}^{i (r)} &=& \oint_{Z_{r}} \frac{dz_s^{(r)}}{2\pi i} 
 \left( \psi^i - \partial X^i\frac{\psi^+}{\partial X^{+}}
 - \frac{1}{2} \psi^i \frac{\psi^+ \partial \psi^+}{(\partial X^+)^2} 
 \right)
   \left(\frac{ i\partial X^+}{p_r^+} \right)^{1/2}
  e^{-i\frac{s}{p^{+}_{r}} X_{L}^{+} }(z_s^{(r)})
~,
\label{eq:DDF-AB}
\end{eqnarray}
and similarly for their anti-holomorphic counterparts
$\tilde{A}^{\tilde{\imath} (r)}_{-\tilde{n}}$
and $\tilde{B}^{\tilde{\imath} (r)}_{-\tilde{s}}$.
Here $X_L^+(z)$ and $X_R^+(\bar{z})$ 
denote the holomorphic and the anti-holomorphic parts of
$X^{+} (z,\bar{z})$. 
$ V_{r}^{\mathrm{DDF}} (Z_r,\bar{Z}_r)$ is the vertex operator
for the DDF state (\ref{eq:DDFstate}), given by
\begin{eqnarray}
&& V_{r}^{\mathrm{DDF}} (Z_r,\bar{Z}_r)
= \epsilon^{(r)}_{\{ij\}}
  A^{i_{1} (r)}_{-n_{1}} \cdots
  \tilde{A}^{\tilde{\imath}_1 (r)}_{-\tilde{n}_{1}} \cdots
  B^{j_{1} (r)}_{-s_{1}} \cdots
  \tilde{B}^{\tilde{\jmath}_1 (r)}_{-\tilde{s}_{1}} \cdots
\nonumber\\
&& \hspace{7em} \times
 e^{ip^{i}_{r} X^{i}} (Z_r,\bar{Z}_r)
 e^{-i p^{+}_{r} X^{-}
     -i \left( p^{-}_{r} - \frac{N_{r}+\tilde{N}_{r}}{2p^{+}_{r}}
         \right)X^{+}}
   (Z_r,\bar{Z}_r)~.
\end{eqnarray}

Since there appears no $\psi^-$ in $V_r^{\mathrm{DDF}}(Z_r,\bar{Z}_r)$,
the expectation values of the operators involving $\psi^+$'s vanish
and thus integrating over $\psi^\pm$ 
yields just the partition function,
in the path integral (\ref{eq:addsector}).
Since $X^-$ appears only in the form $e^{-ip_r^{+}X^-}$, 
it is also easy to carry out the path integral over $X^\pm$
and we obtain
\begin{eqnarray}
&&\int [dX^\pm]e^{-S^{X}_{\pm}}
\prod_{r=1}^N
\left[
 e^{-i\frac{n_1}{p^{+}_{r}} X_{L}^{+} }(z_{n_1}^{(r)})
\cdots 
 \left(\frac{ i\partial X^+}{p_r^+} \right)^{1/2}
  e^{-i\frac{s_1}{p^{+}_{r}} X_{L}^{+} }(z_{s_1}^{(r)})
\cdots
\right.
\nonumber\\
&& \qquad  \left. \times
e^{-i\frac{\tilde{n}_1}{p^{+}_{r}} X_{R}^{+} }
     (\bar{z}_{\tilde{n}_1}^{(r)})
\cdots 
 \left(\frac{ i\bar{\partial} X^+}{p_r^+} \right)^{1/2}
  e^{-i\frac{\tilde{s}_1}{p^{+}_{r}} X_{R}^{+} }
      (\bar{z}_{\tilde{s}_1}^{(r)})
\cdots
e^{-i p^{+}_{r} X^{-}
     -i \left( p^{-}_{r} - \frac{N_{r}+\tilde{N}_{r}}{2p^{+}_{r}}
         \right)X^{+}}
   (Z_r,\bar{Z}_r)
\right]
\nonumber\\
&& \sim
 (2\pi)^2 \delta\left(\sum_{r=1}^N p_r^+\right)
\delta\left(\sum_{r=1}^N p_r^-\right)
\prod_{r=1}^N
\left[  \left(w_{n_1}^{(r)} \right)^{-n_1}
\cdots \left(\frac{\partial w_{s_1}^{(r)}}
            {\partial z_{s_1}^{(r)}}\right)^{\frac{1}{2}}
\left(w_{s_1}^{(r)} \right)^{-s_1-\frac{1}{2}} 
\cdots\right.
\nonumber\\
&& \quad  \times
\left.
\left(\bar{w}_{\tilde{n}_1}^{(r)} \right)^{-\tilde{n}_1}
\cdots \left(\frac{\partial \bar{w}_{\tilde{s}_1}^{(r)}}
        {\partial \bar{z}_{\tilde{s}_1}^{(r)}}\right)^{\frac{1}{2}}
\left(\bar{w}_{\tilde{s}_1}^{(r)} \right)
           ^{-\tilde{s}_1-\frac{1}{2}} 
\cdots
e^{\left( p^{i}_{r} p^{i}_{r}
             -1  \right)
      \mathrm{Re} \bar{N}^{rr}_{00} 
             }
e^{-p_r^- \tau_0^{(r)}}
\right]~,
\label{eq:integralxpm}
\end{eqnarray}
where
$S^{X}_{\pm}$ is the $X^{\pm}$ part of the worldsheet action
$S_{\pm}$,
and $w_{n}^{(r)}$ denotes the coordinate on the
unit disk $w_r$ defined in eq.(\ref{eq:rho-wr}) 
which corresponds to $z_n^{(r)}$.
Here we have used the level-matching condition and 
the on-shell condition (\ref{eq:onshell}).
By using eq.(\ref{eq:integralxpm}) and the relation
\begin{equation}
e^{ip^{i}_{r}X^{i}} (Z_{r},\bar{Z}_{r})
= e^{- p^{i}_{r}p^{i}_{r} 
       \mathrm{Re} \bar{N}^{rr}_{00}  }
     e^{ip^{i}_{r}X^{i (r)}} (w_{r}=0,\bar{w}_{r}=0)~,
\end{equation}
we obtain eq.(\ref{eq:addsector}).

With the ghosts and the longitudinal coordinates
thus introduced, we have the expression
\begin{eqnarray}
\lefteqn{
F (\mathcal{T}_{I},\bar{\mathcal{T}}_{I})
\sim \int [dXd\psi db dc d\beta d\gamma d\phi^\mathrm{ext}] 
e^{-S }|\mathcal{D}|^{2c^\mathrm{ext} }
     \left( \lim_{z,\bar{z} \rightarrow \infty}
    \frac{1}{|z|^{4}} c(z) \tilde{c} (\bar{z})
    \right)
     \left| \sum_{s=1}^{N} \alpha_{s} Z_{s} \right|^{2}
}
\nonumber\\
&& \qquad
   \times \prod_{r=1}^{N} \left[
     c \tilde{c} 
     e^{-\phi}  e^{-\tilde{\phi} } 
     V^{\mathrm{DDF}}_{r} (Z_{r},\bar{Z}_{r})
     \right]
 \prod_{I} \left[
    \frac{b(z_{I})}{\partial^{2} \rho (z_{I})}
    \frac{\tilde{b} (\bar{z}_{I} )}
         {\bar{\partial}^{2} \bar{\rho} (\bar{z}_{I})}
 T_F^\mathrm{LC}  e^{ \phi} (z_{I})
 \tilde{T}_F^\mathrm{LC}  
      e^{\tilde{\phi} }(\bar{z}_{I})
   \right],~~~~~
\label{eq:F-cov2}
\end{eqnarray}
where $S$ is the total worldsheet action 
$S =S^{\mathrm{LC} }+ S_{\pm} + S_{\mathrm{gh}}$.

\subsubsection*{Picture changing operator}

In the following, we will show that
$T_F^\mathrm{LC}$ in eq.(\ref{eq:F-cov2}) can be 
rewritten by using the picture changing operator in ten dimensions. 
Let us introduce a fermionic charge $Q$
defined as
\begin{equation}
Q= \frac{1}{2}\oint \frac{dz}{2\pi i} \partial \rho (z)
 \Big( c(z) \left(2i\partial X^{+} (z) - \partial \rho(z)
  \right)
       + \eta (z) e^{\phi }(z)  \psi^{+} (z)
 \Big)~.
 \label{eq:Q-charge}
\end{equation}
We notice that this charge is nilpotent $Q^{2} = 0$, and
because of the relations
\begin{eqnarray}
&&
\left[ Q\, , \,c \tilde{c} e ^{-\phi }e^{-\tilde{\phi}}
V_r^{\mathrm{DDF}}(Z_r,\bar{Z}_r) \right]=0~,
\quad
\left[ Q \, , \,  T^{\mathrm{LC}}_{F} e^{\phi} (z_{I})
\right]=0~,
\nonumber\\
&& 
\left\{ Q \, , \, \frac{b(z_{I})}{\partial^{2} \rho (z_{I})}
\right\}=0~,
\quad 
\left\{ Q \, , \, c(\infty) \right\}=0~,
\end{eqnarray}
$Q$ commutes with the operators which appear on the right hand side of 
 eq.(\ref{eq:F-cov2}).
Therefore, we can replace 
$T^{\mathrm{LC}}_{F} e^{\phi} (z_{I})$ in eq.(\ref{eq:F-cov2})
by
\begin{eqnarray}
X'(z_{I})
 &\equiv&
 T^{\mathrm{LC}}_{F} e^{\phi} (z_{I})
  - \frac{1}{\partial^{2} \rho (z_{I})}
    \left\{ Q \, , \, i\partial X^{-} \partial \xi (z_{I})
                + \frac{1}{2} \partial b 
                   e^{\phi} \psi^{-} (z_{I})
   \right\}
\nonumber\\
&=& \left(  T^{\mathrm{LC}}_{F} e^{\phi}
   + c \partial \xi
   + \frac{i}{2}
      \left(\psi^{+}\partial X^{-} + \psi^{-} \partial X^{+}
      \right)
      e^{\phi}
   + \frac{1}{4} \partial b \eta e^{2\phi}
  \right) (z_{I})~,
\label{eq:Xprime}
\end{eqnarray}
without changing $F (\mathcal{T}_{I},\bar{\mathcal{T}}_{I})$. 
$X'(z_{I})$ thus defined satisfies
\begin{equation}
b (z_{I}) X' (z_{I})
= b(z_{I}) 
  \left(
  \{Q_{\mathrm{B}}\, , \, \xi (z_{I})\}
    + T_F^\mathrm{ext} e^\phi (z_{I})
  \right)~,
\end{equation}
where $Q_{\mathrm{B}}$ denotes the BRST charge
of the covariant NSR superstring
theory in the ten dimensional space-time, and
$\{Q_{\mathrm{B}} \, , \xi (z_{I})\}$
is the picture changing operator,
which takes the form
\begin{equation}
\{ Q_{\mathrm{B}}\, , \, \xi (z) \}
 = -\frac{1}{2} i \psi^{\mu} \partial X_{\mu} e^{\phi} (z)
     + c \partial \xi (z)
     + \frac{1}{4} \partial b \eta e^{2\phi} (z)
    + \frac{1}{4} b \left( 2 \partial \eta e^{2\phi}
                           + \eta (\partial e^{2\phi})
                    \right) (z)~.
\end{equation} 

Consequently, we find that $T^{\mathrm{LC}}_{F} e^{\phi} (z_{I})$
in the path integral (\ref{eq:F-cov2}) can be replaced by
$\{Q_{\mathrm{B}}\, , \, \xi (z_{I})\}
 + T_F^\mathrm{ext} e^\phi (z_{I})$, and we obtain
\begin{eqnarray}
&&
F (\mathcal{T}_{I},\bar{\mathcal{T}}_{I})
\nonumber\\
&&
\sim \int [dXd\psi db dc d\beta d\gamma d\phi^\mathrm{ext}] 
e^{-S }
\nonumber\\
&&\times |
\mathcal{D}|^{2c^\mathrm{ext} }
     \left( \lim_{z,\bar{z} \rightarrow \infty}
    \frac{1}{|z|^{4}} c(z) \tilde{c} (\bar{z})
    \right)
     \left| \sum_{s=1}^{N} \alpha_{s} Z_{s} \right|^{2}
 \prod_{r=1}^{N} \left[
     c \tilde{c} e^{-\phi} e^{-\tilde{\phi} } 
     V^{\mathrm{DDF}}_{r} (Z_{r},\bar{Z}_{r})
     \right]
\nonumber\\
&& 
   \times \prod_{I} \left[
    \frac{b(z_{I})}{\partial^{2} \rho (z_{I})}
    \frac{\tilde{b} (\bar{z}_{I} )}
         {\bar{\partial}^{2} \bar{\rho} (\bar{z}_{I})}
    \left( \{Q_{\mathrm{B}}\, , \, \xi (z_{I}) \}
           +T_F^\mathrm{ext}  e^{ \phi} (z_{I})
    \right)  
    \left(
    \{ \tilde{Q}_{\mathrm{B}}\, , \, \tilde{\xi}(\bar{z}_{I}) \}
    + \tilde{T}_F^\mathrm{ext}  
      e^{\tilde{\phi} }(\bar{z}_{I})
    \right)
   \right].~~~~~
\label{eq:F-cov4}
\end{eqnarray}

\section{Regularization of the Four-point Tree Amplitudes}
\label{sec:4pt}

In this section, we restrict our attention to the $N=4$ case.
We define the amplitudes for large negative $c^\mathrm{ext}$'s and 
analytically continue them to $c^\mathrm{ext}=0$. 
We would like to compare our amplitudes with the ones 
obtained in the first-quantized formalism.

\begin{figure}[htbp]
\begin{center}
	\includegraphics[width=35em]{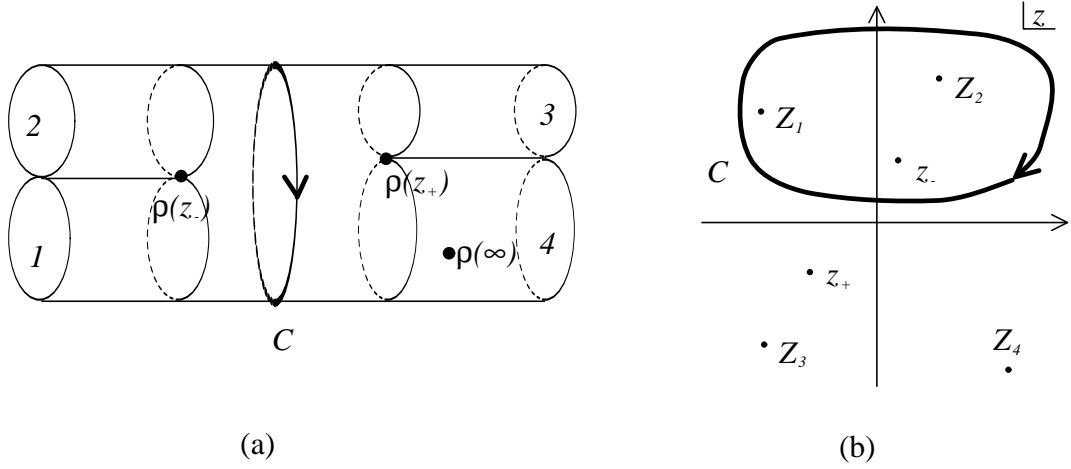}
	\caption{The contour $C$ on the $\rho$-plane
                 of the integral
                 (\ref{eq:F-cov5}) for the $N=4$ case
                is depicted in (a).
                On the $z$-plane, it is described in (b).}
	\label{fig:4pt}
\end{center}
\end{figure}

By using the relation
\begin{equation}
\frac{b(z_{I})}{\partial^{2} \rho (z_{I})}
 = \oint_{z_{I}} \frac{dz}{2\pi i}
   \frac{b(z)}{\partial \rho (z)}~,
\end{equation}
and deforming the contour of this integral,
one can simplify eq.(\ref{eq:F-cov4}) for $N=4$:
\begin{eqnarray}
F (\mathcal{T},\bar{\mathcal{T}})
&\sim& \int [dXd\psi db dc d\beta d\gamma d\phi^\mathrm{ext}] 
\,  e^{-S} \, |\mathcal{D}|^{2c^\mathrm{ext} }
\oint_C\frac{dz}{2\pi i}\frac{b}{\partial\rho}
\oint_C\frac{d\bar{z}}{2\pi i}\frac{\tilde{b}}{\bar{\partial}\bar{\rho}}
\nonumber\\
&& \!\!
   \times \prod_{r=1}^{4} \left[
     c(Z_{r}) \tilde{c} (\bar{Z}_{r})
     e^{-\phi } (Z_{r})  e^{-\tilde{\phi} }(\bar{Z}_{r}) 
     V^{\mathrm{DDF}}_{r} (Z_{r},\bar{Z}_{r})
     \right]
\nonumber\\
&& \!\!
   \times \prod_{I=\pm} \left[
    \left( \{Q_{\mathrm{B}}\, , \, \xi (z_{I}) \}
             +T_F^\mathrm{ext} e^{\phi} (z_{I})
    \right)  
    \left(
    \{ \tilde{Q}_{\mathrm{B}}\, , \, \tilde{\xi}(\bar{z}_{I}) \}
    + \tilde{T}_F^\mathrm{ext} 
       e^{\tilde{\phi} } (\bar{z}_{I})
    \right)
   \right]~.
\label{eq:F-cov5}
\end{eqnarray}
Here the integration contour $C$ is depicted in Fig.~\ref{fig:4pt}.
$z_{\pm}$ denote the two interaction points defined by
$z_{-} \equiv z^{(1)}_{I} = z^{(2)}_{I}$,
$z_{+} \equiv z^{(3)}_{I} = z^{(4)}_{I}$,
and thus the one moduli parameter $\mathcal{T}$ becomes
$\mathcal{T} = \rho (z_{+}) - \rho (z_{-})$.

Now it is easy to see that
the integrand $F (\mathcal{T},\bar{\mathcal{T}})$
is expressed as 
\begin{equation}
F(\mathcal{T},\bar{\mathcal{T}})
\sim \left| 
\mathcal{F}_\mathrm{10D} (\mathcal{T}) 
-
\frac{c^\mathrm{ext}}{6(z_+-z_-)^3}\mathcal{F}' (\mathcal{T})
\right|^{2}
  \left| \mathcal{D} (\mathcal{T}) \right|^{2c^{\mathrm{ext}}}~,
\end{equation}
where
\begin{eqnarray}
\mathcal{F}_\mathrm{10D} &=& \left\langle
\oint_C\frac{dz}{2\pi i}\frac{b}{\partial\rho}
  \prod_{r=1}^{4} V_{r}^{(-1)} (Z_{r})
  \prod_{I=\pm}
     \left[ 
            \{ Q_{\mathrm{B}},\xi (z_{I}) \}
    \right]   \right\rangle~,
\nonumber\\
\mathcal{F}' &=& \left\langle
\oint_C\frac{dz}{2\pi i}\frac{b}{\partial\rho}
  \prod_{r=1}^{4} V_{r}^{(-1)} (Z_{r})
  \prod_{I=\pm}
             e^\phi (z_{I})
     \right\rangle~.
\label{eq:correlationF}
\end{eqnarray}
Here $\langle \cdots \rangle$
denotes the correlation function of the system
$(X^{\mu},\psi^{\mu}; b,c;\beta,\gamma)$,
and $V_{r}^{(-1)} (Z_{r})$ is defined by using
the holomorphic part $V^{\mathrm{DDF}}_{rL} (Z_{r})$ of
$V^{\mathrm{DDF}}_{r} (Z_{r},\bar{Z}_{r})$ as
\begin{equation}
V^{(-1)}_{r} (Z_{r})= c e^{-\phi} 
V^{\mathrm{DDF}}_{rL} (Z_{r})~.
\end{equation}

$\mathcal{F}_\mathrm{10D}$ is almost the first-quantized result,
except that the picture changing operators are inserted 
at $z_\pm$. 
These can be moved in the usual manner~\cite{Friedan:1985ge} 
up to total derivative terms with respect to the moduli parameter. 
As a result we obtain
\begin{equation}
\mathcal{F}_\mathrm{10D}
=
\mathcal{F}_\mathrm{fin}+\partial_\mathcal{T}\mathcal{G}~,
\end{equation}
where
\begin{equation}
\mathcal{F}_{\mathrm{fin}}
 = \left\langle
   \oint_{C} \frac{dz}{2\pi i } \frac{b(z)}{\partial \rho (z)}
   V^{(0)}_{1} (Z_{1}) V^{(0)}_{2}(Z_{2})
   V^{(-1)}_{3} (Z_{3}) V^{(-1)}_{4} (Z_{4})
  \right\rangle~,
\end{equation}
and $\mathcal{G}$ can be expressed by a correlation function 
in the large Hilbert space. 
$V^{(0)}_{r} (Z_{r})$ is defined as
\begin{equation}
  V^{(0)}_{r} (Z_{r})
 = \oint_{Z_{r}} \frac{dz}{2\pi i} j_{\mathrm{B}}(z)
  \xi (Z_{r}) V^{(-1)}_{r} (Z_{r})~,
\end{equation}
where $j_{\mathrm{B}}(z)$ denotes the BRST current.

Collecting the results obtained above,
we find that the four-point tree amplitudes
become
\begin{equation}
\mathcal{A} \sim - g^2 C_4 \int \frac{d^{2} \mathcal{T} }{4\pi}
 \left| \mathcal{F}_{\mathrm{fin}}
        + \partial_{\mathcal{T}} \mathcal{G}
        -\frac{c^\mathrm{ext}}{6(z_+-z_-)^3}
        \mathcal{F}'
         \right|^2
 |\mathcal{D}|^{2c^{\mathrm{ext}}}~.
\label{eq:amplitude-4pt}
\end{equation}

\subsubsection*{Possible divergences}

We would like to study the possible divergences in the amplitudes 
when we take $c^\mathrm{ext}\to 0$. 
In the subsequent computations, we take 
\begin{equation}
Z_{1}=1~, \qquad Z_{2}=0~,\qquad Z_{3} = Z~, \qquad
Z_{4}=\infty~,
\end{equation}
and regard $Z$ as a function of $\mathcal{T}$.
In this gauge, the interaction points $z_{\pm}$ become
\begin{equation}
z_{\pm}
 = - \frac{(\alpha_{1}+\alpha_{2})Z+\alpha_{2}+\alpha_{3}}
          {2 \alpha_{4}}
   \pm \frac{1}{2}
       \frac{\alpha_{1}+\alpha_{2}}{\alpha_{4}}
   \sqrt{\left(
          Z + \frac{\alpha_{2}+\alpha_{3}}{\alpha_{1}+\alpha_{2}}
         \right)^{2}
         + \frac{4\alpha_{2} \alpha_{4}}
                 {(\alpha_{1}+\alpha_{2})^{2}} Z }~.
\end{equation}
We write
\begin{equation}
z_{+} - z_{-}
 = \frac{\alpha_{1}+\alpha_{2}}{\alpha_{4}}
   \sqrt{(Z-X)(Z-Y)}~.
\label{eq:XY}
\end{equation}
The Jacobian for the change of the variables from $\mathcal{T}$
into $Z$ becomes
\begin{equation}
\frac{\partial \mathcal{T}}{\partial Z}
=\frac{\alpha_{4} (z_{+} - z_{-})}{Z (Z-1)}~. 
\end{equation}
The amplitude (\ref{eq:amplitude-4pt})
turns out to be
\begin{eqnarray}
\mathcal{A}
 &\sim& - \frac{g^2 C_4}{4\pi} \int d^{2} Z
   \left| \frac{\partial \mathcal{T}}{\partial Z}
          \left(\mathcal{F}_{\mathrm{fin}}
          -\frac{c^{\mathrm{ext}}}{6(z_+-z_-)^3}\mathcal{F}'
          \right)
           + \partial_{Z} \mathcal{G} \right|^2
  |\mathcal{D}|^{2 c^{\mathrm{ext}}}
\nonumber\\
&=& - \frac{g^2 C_4}{4\pi} \int d^{2}Z 
 \left| \frac{\partial \mathcal{T}}{\partial Z} \right|^{2}
 \left| \mathcal{F}_{\mathrm{fin}} \right|^{2}
 \left|\mathcal{D} \right|^{2 c^{\mathrm{ext}}}
\nonumber\\
&& {}+ c^{\mathrm{ext}}\frac{g^2 C_4}{4\pi}  \int d^{2}Z \left[
   \frac{\partial \mathcal{T}}{\partial Z}
          \mathcal{F}_{\mathrm{fin}}
   \left(
   \frac{1}{6(\bar{z}_+-\bar{z}_-)^3}
   \frac{\partial \bar{\mathcal{T}}}{\partial \bar{Z}}
    \bar{\mathcal{F}}'+
   \bar{\mathcal{G}}
   \frac{\partial_{\bar{Z}} \bar{\mathcal{D}}}
        {\bar{\mathcal{D}}}
   \right)
   \left| \mathcal{D} \right|^{2c^{\mathrm{ext}}}
  +
   \mathrm{c.c.}
  \right]
\nonumber\\
&& {}- (c^{\mathrm{ext}})^{2} \frac{g^2 C_4}{4\pi} 
  \int d^{2}Z  
  \left|
   \frac{1}{6(z_+-z_-)^3}
   \frac{\partial \mathcal{T}}{\partial Z}
          \mathcal{F}'
   +
   \mathcal{G}
   \frac{\partial_{Z} \mathcal{D}}
        {\mathcal{D}}
   \right|^2
   \left| \mathcal{D} \right|^{2c^{\mathrm{ext}}}~.
\label{eq:amplitude-4pt2}
\end{eqnarray}
Here we have taken $c^\mathrm{ext}$ so that no surface terms
appear in the integration by parts.

Let us analytically continue $c^{\mathrm{ext}}$ to $0$.
First, we consider the first term on the right hand side
of eq.(\ref{eq:amplitude-4pt2}).
Using
\begin{eqnarray}
\frac{\partial \mathcal{T}}{\partial Z}
\mathcal{F}_{\mathrm{fin}}
 &=& 
 \left(
 \frac{\partial \rho (z_+)}{\partial Z}
 -
 \frac{\partial \rho (z_-)}{\partial Z}
 \right)
 \left\langle
   \oint_{C} \frac{dz}{2\pi i } \frac{b(z)}{\partial \rho (z)}
   V^{(0)}_{1} (Z_{1}) V^{(0)}_{2}(Z_{2})
   V^{(-1)}_{3} (Z) V^{(-1)}_{4} (Z_{4})
  \right\rangle
\nonumber\\
&=&
 \Bigg\langle
 \left(
   \oint_{C} \frac{dz}{2\pi i } 
   \frac{\partial_Z\rho (z)-\partial_Z\rho (z_-)}
   {\partial \rho (z)}b(z)
   -
   \oint_{C} \frac{dz}{2\pi i } 
   \frac{\partial_Z\rho (z)-\partial_Z\rho (z_+)}
   {\partial \rho (z)}b(z)
  \right)
  \nonumber\\
  & &\hspace{1cm}
  \times
   V^{(0)}_{1} (Z_1) V^{(0)}_{2}(Z_{2})
   V^{(-1)}_{3} (Z) V^{(-1)}_{4} (Z_{4})
  \Bigg\rangle~,
\end{eqnarray}
and deforming the contour for the integration of $b$, 
we obtain
\begin{equation}
\frac{\partial \mathcal{T}}{\partial Z}
\mathcal{F}_{\mathrm{fin}}
=
 \left\langle
   V^{(0)}_{1} (Z_{1}) V^{(0)}_{2} (Z_{2})
   \left(e^{-\phi} V^{\mathrm{DDF}}_{3L}\right) (Z)
    V^{(-1)}_{4} (Z_{4})
  \right\rangle~,
\end{equation}
and find that $\frac{\partial \mathcal{T}}{\partial Z}
\mathcal{F}_{\mathrm{fin}}$ involves no divergences.
Therefore, 
the first term on the right hand side
of eq.(\ref{eq:amplitude-4pt2}) 
turns out to be
\begin{equation}
-\frac{g^2 C_4}{4\pi}\int d^2Z
\left| \frac{\partial \mathcal{T}}{\partial Z} \right|^{2}
 \left| \mathcal{F}_{\mathrm{fin}} \right|^{2}~,
\end{equation}
without any problem
and coincides with the amplitude of the first-quantized
formalism.

In the following, we will see that 
the second and the third terms are vanishing as
$c^{\mathrm{ext}} \rightarrow 0$.
Since these terms are
proportional to $c^{\mathrm{ext}}$ and $(c^{\mathrm{ext}})^{2}$,
it is sufficient to show 
that the integrals in these terms do not give rise to dangerous
divergences as $c^{\mathrm{ext}} \rightarrow 0$.
The regions in the moduli space where these integral
can be singular are as follows:
$(i)$ $Z \rightarrow Z_{1},Z_{2},Z_{4}$;
$(ii)$ $Z \rightarrow z_{\pm}$;
$(iii)$ $Z \rightarrow X,Y$,
where $X$ and $Y$ are given in eq.(\ref{eq:XY}),
i.e.\ $z_{+} \rightarrow z_{-}$.
The possible singularity in region~$(i)$ can be
dealt with by the analytic continuation of the
external momenta $p_{r}$.
We therefore need not worry about
the singularity in this case.
Because of eq.(\ref{eq:Zr-Zs}), case~$(ii)$ is reduced to
case~$(i)$ and need not be worried about, either.
The third case is what we should study carefully.
In the following,
we will show that  the second and the third
terms on the right hand side of eq.(\ref{eq:amplitude-4pt2})
do not provide dangerous contributions
in region $(iii)$ and hence
vanish as $c^{\mathrm{ext}} \rightarrow 0$.

Let us begin by considering the second term.
We divide the integration region into three parts as
\begin{equation}
\int d^2Z
=
\left(
\int_{|Z-X| < \epsilon}
+
\int_{|Z-Y| < \epsilon}
+
\int_{|Z-X| > \epsilon \ \mathrm{and} \ |Z-Y| > \epsilon }
\right)
d^2Z~.
\end{equation}
The possible divergences come from the first and the second regions. 
Let us focus on the  region $|Z-X| < \epsilon$.
One can find that
$\frac{\partial \mathcal{T}}{\partial Z}
 \mathcal{F}_{\mathrm{fin}}$
can be expanded by the Taylor series, and  
$\frac{\partial \mathcal{T}}{\partial Z}
 \frac{\mathcal{F}'}{\left(z_+-z_-\right)^3}$ and
$\mathcal{G}\frac{\partial_Z\mathcal{D}}{\mathcal{D}}$
can be expanded by the Laurent series 
in $(Z-X)$, respectively.
Taking eq.(\ref{eq:D}) into account, we find that
the integral of each term in the expansion behaves
around $Z \sim X$ as
\begin{eqnarray}
&& \int_{|Z-X| < \epsilon} d^{2} Z 
   (Z-X)^{ n - \frac{c^{\mathrm{ext} }}{96}}
   (\bar{Z} - \bar{X})^{m -  \frac{ c^{\mathrm{ext}} }{96}}
\nonumber\\ 
&& \quad \sim \int^{\epsilon}_{0} dr
           \int^{2\pi}_{0} d\theta
           \, r^{ (n+ m)+1-\frac{ c^{\mathrm{ext}} }{48}}
   e^{i \theta (n-m)}
= 2\pi \delta_{n,m}\int^{\epsilon}_{0} dr
           \, r^{ 2n+1-\frac{ c^{\mathrm{ext}} }{48}}~.
\end{eqnarray}
Here $n$ and $m$ are integers
and at least one of them is nonnegative.
This contribution is finite in the limit 
$c^{\mathrm{ext}} \rightarrow 0$. 
The same argument can be applied to the region 
$|Z-Y|<\epsilon$. 
Therefore, the second term on the right hand side
of eq.(\ref{eq:amplitude-4pt2}) vanishes
in $c^{\mathrm{ext}} \rightarrow 0$.

Let us turn to the integral of the third term
on the right hand side of eq.(\ref{eq:amplitude-4pt2}).
Due to the same discussion given above,
the singular contributions are of the form
\begin{equation}
\int_{|Z-X| <\epsilon} d^{2} Z (Z-X)^{-1-\frac{ c^{\mathrm{ext}}}{96}}
         (\bar{Z} - \bar{X})^{-1-\frac{ c^{\mathrm{ext}}}{96}}
 = \mathcal{O} \left( \frac{1}{c^{\mathrm{ext}}} \right)~,
\end{equation}
around $Z \sim X$ for example, and 
we find that the possible divergence is at 
most $\mathcal{O}\left(\frac{1}{c^{\mathrm{ext}}}\right)$.
Therefore the third term on the right hand side of
eq.(\ref{eq:amplitude-4pt2}),
obtained from such integrals multiplied
by $(c^{\mathrm{ext}})^{2}$,  is  vanishing
as $c^{\mathrm{ext}} \rightarrow 0$.

Thus
we find that the regularization prescription
works in the four-point amplitudes 
without counterterms
and the results of the first-quantized formalism 
are reproduced.

\section{Discussions}
\label{sec:discussion}

In this paper, we have proposed a dimensional regularization
scheme to regularize the divergences caused by the
colliding transverse supercurrents inserted 
at the interaction points of the three-string vertices
of the light-cone gauge superstring field theory. 
We have investigated the tree amplitudes and shown that the
divergences originating in the colliding supercurrents
are actually regularized.
We have explicitly studied the four-point amplitudes of
the NS-NS closed strings,
and shown that the usual results are reproduced without
any counterterms in this case.
Although the amplitudes without the Ramond strings have been considered,
we can treat the amplitudes including them in a similar way .

There are many problems that remain to be studied.
We should go on and examine the tree amplitudes with $N \geq 5$
and then the higher-loop amplitudes, and
see if we need counterterms to reproduce the desired results. 
In these cases, there are situations in which more than two
interaction points approach one another, which makes 
the integrations over the moduli spaces more complicated. 

If it is difficult to treat the amplitudes directly, 
it may be better to pursue other approaches. 
In the dimensional regularization,
we shift the space-time dimension away from the critical
dimension.
The light-cone gauge string field theory in such a space-time 
is not Lorentz invariant. 
However, since the light-cone gauge string field theory is 
a gauge fixed theory, it will be possible to find 
a Lorentz noninvariant but gauge invariant theory
whose gauge fixed version coincides with 
such a noncritical string theory. 
If such a theory exists, 
the form of the possible counterterms will be 
severely restricted. 
The progress in this direction will be reported
elsewhere.

\section*{Acknowledgements}

We would like to thank Y. Satoh for discussions. 
This work was supported in part by 
Grant-in-Aid for Scientific Research (C) (20540247)
and
Grant-in-Aid for Young Scientists (B) (19740164) from
the Ministry of Education, Culture, Sports, Science and
Technology (MEXT).


\newpage
\appendix

\section{Action of String Field Theory}
\label{sec:action}

We present the light-cone gauge string field theory
of NSR closed superstrings  containing the extra sector
explained in section \ref{sec:lc-amp}.
Since we deal with the NS-NS sector in this paper, 
we consider the part involving only this sector. 
We define the action for the NS-NS sector as
\begin{eqnarray}
S &=&  \int dt \Biggl[ \frac{1}{2}
  \int d1 d2 \, \langle R(1,2)|\Phi \rangle_{1}
  \left( i\frac{\partial}{\partial t}
          - \frac{L_{0}^{\mathrm{LC}(2)}
       +\tilde{L}_{0}^{\mathrm{LC}(2)} -\frac{c}{12} }{\alpha_2}
        \right)
  |\Phi \rangle_{2}
\nonumber\\
 && \hspace{3em}
   {}+ \frac{2g}{3}   
 \int d1d2d3 \, \langle V_{3} (1,2,3) |
  \Phi \rangle_{1} |\Phi \rangle_{2}
  |\Phi\rangle_3
  \Biggr]~,
\label{eq:sftaction}
\end{eqnarray}
where $t$ denotes the proper time,
$c$ is the central charge given in eq.(\ref{eq:center}),
$g$ is the coupling constant,
$\alpha_{r}=2p^{+}_{r}$ denotes the string-length parameter of
the $r$-th string,
and $dr$ is the integration measure for
the momentum zero modes of the $r$-th string,
\begin{equation}
dr = 
   \frac{\alpha_r d\alpha_{r}}{4\pi}
   \frac{d^{8}p_{r}}{(2\pi)^{8}}~.
\label{eq:measure}
\end{equation}
The string field $\Phi$ is taken to be Grassmann even
and subject to
\begin{eqnarray}
&&\int_0^{2\pi} \frac{d\theta}{2\pi } 
e^{i\theta(L_0^{\mathrm{LC}} -\tilde{L}_0^{\mathrm{LC}})}
 |\Phi \rangle
 = |\Phi \rangle~,
\quad
\mathcal{P}^{\mathrm{GSO}} |\Phi \rangle
 = |\Phi \rangle~,
\label{eq:projectedfield}
\end{eqnarray}
where $L_0^{\mathrm{LC}}$ is the transverse Virasoro zero mode
of the worldsheet theory
of the light-cone gauge NSR superstrings containing
the extra sector
and $\mathcal{P}^{\mathrm{GSO}}$ denotes 
usual GSO projection operator for the light-cone gauge superstrings.
The reflector $\langle R (1,2)|$ and the three-string vertex 
$\langle V_{3} (1,2,3)|$
are defined as
\begin{eqnarray}
\langle R (1,2)|
 &=& \delta (1,2) \, 
 \, {}_{2} \langle 0 |\,{}_{1} \langle 0 |\,
 e^{ -\sum_{n=1}^{\infty}
      \frac{1}{n} \alpha^{i(1)}_{n} \alpha^{i(2)}_{n}
 + i\sum_{k>0} \psi^{i(1)}_{k} \psi^{i(2)}_{k} + 
 \mathrm{extra\ sector}
 +
 \mathrm{c.c.}}~,
\nonumber\\
\langle V_{3}(1,2,3)| &=& 4\pi
\delta\left(\sum_{r=1}^N \alpha_r \right)
\langle V_{3}^{\mathrm{LPP}}(1,2,3)|
   e^{-\Gamma^{\mathrm{LC[3]}}}
     \left( \partial^{2} \rho (z_{I})
                  \bar{\partial}^{2} \bar{\rho}(\bar{z}_{I})
           \right)^{-\frac{3}{4}}
          T^{\mathrm{LC}}_{F} (z_{I})
          \tilde{T}_{F}^{\mathrm{LC}} (\bar{z}_{I})~,
\nonumber\\
\delta (1,2) &=& (2\pi)^{8} \delta^{8} (p_{1}+p_{2})
  \frac{4\pi}{\alpha_{1}}
  \delta (\alpha_{1}+\alpha_{2})~,
\nonumber\\
 e^{-\Gamma^{\mathrm{LC[3]}}}
   &=& 
   \mathrm{sgn}\left(\alpha_1\alpha_2 \alpha_3\right)
   \left|
     \frac{e^{-2\hat{\tau}_0\sum_{r=1}^3 
     \frac{1}{\alpha_r}}}{\alpha_1\alpha_2\alpha_3}\right|
     ^{\frac{c}{24}}~,
  \qquad \hat{\tau}_0= \sum_{r=1}^3\alpha_r\ln |\alpha_r|~,
\label{eq:exp-Gamma-3}
\end{eqnarray}
where $\mathrm{sgn}(x)$ denotes the sign function,
$\langle V_{3}^{\mathrm{LPP}}(1,2,3)|$
denotes the LPP vertex for three NS-NS closed strings
with extra sector and
$T^{\mathrm{LC}}_{F} $ is transverse supercurrent
of the worldsheet theory of the light-cone gauge NSR superstring
containing the extra CFT,
\begin{equation}
T_{F}^{\mathrm{LC}}
 = T^{(X,\psi)\mathrm{LC}}_{F}+T^{\mathrm{ext}}_{F}~.
\label{eq:Tf}
\end{equation}
The Mandelstam mapping $\rho(z)$ and the interaction point $z_I$
are defined in the next appendix.

\section{Mandelstam Mapping}
\label{sec:Mandelstam}

\begin{figure}[htbp]
\begin{center}
	\includegraphics[width=20em]{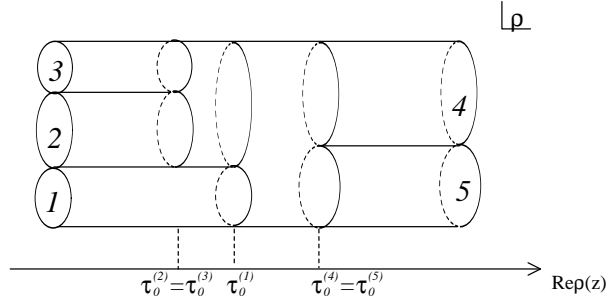}
	\caption{A typical $N$-string tree diagram with $N=5$.}
	\label{fig:mandelstam1}
\end{center}
\end{figure}

We introduce the complex coordinate $\rho$ on the $N$-string
tree diagram with the joining-splitting type of interaction
described in Fig.~\ref{fig:mandelstam1}.
Each portion on the $\rho$-plane corresponding to
the $r$-th external string $(r=1,\ldots,N)$ is identified with
the unit disk $|w_{r}| \leq 1$ of the $r$-th string by the relation
\begin{equation}
\rho = \alpha_{r} \ln w_{r} + \tau^{(r)}_{0} + i \beta_{r}~,
\label{eq:rho-wr}
\end{equation}
where $\tau^{(r)}_{0}+ i\beta_{r}$ is the coordinate
on the $\rho$-plane at which the $r$-th string
interacts (Fig.~\ref{fig:mandelstam1}).

The $N$-string tree diagram can be mapped to the
complex $z$-plane with $N$ punctures by the Mandelstam mapping,
\begin{equation}
\rho (z) = \sum_{r=1}^{N} \alpha_{r} \ln (z-Z_{r})~,
\qquad \sum_{r=1}^{N} \alpha_{r} =0~,
\label{eq:Mandelstam}
\end{equation}
where the point $z=Z_{r}$ $(r=1,\ldots,N)$ is the puncture
corresponding to the origin of the unit disk $w_{r}=0$.
The interaction points $z_{I}$ are determined by
$\frac{\partial \rho}{\partial z} (z_{I})=0$.
These are related to $\tau^{(r)}_{0}$ and $\beta_{r}$
introduced above by the equation
\begin{equation}
\tau^{(r)}_{0} + i \beta_{r}
= \rho (z^{(r)}_{I})~,
\label{eq:rho-zIr}
\end{equation}
where $z^{(r)}_{I}$ is the interaction point
on the $z$-plane
at which the $r$-th external string interacts.

{}From eq.(\ref{eq:Mandelstam}) and the definition of the
interaction points $z_{I}$, we obtain
\begin{eqnarray}
&&
\partial^{2} \rho (z_{I})
 = \left( \sum_{s=1}^{N} \alpha_{s} Z_{s} \right)
   \frac{\prod_{J \neq I} (z_{I} - z_{J})}
        {\prod_{r=1}^{N} (z_{I} - Z_{r}) }~,
\label{eq:deldelrho}
\\
&&
\alpha_{r} \prod_{s \neq r} (Z_{r} - Z_{s})
 = \left( \sum_{s=1}^{N} \alpha_{s} Z_{s} \right)
   \prod_{I} (Z_{r} - z_{I})~.
\label{eq:Zr-Zs}
\end{eqnarray}

\section{Derivation of $e^{-\Gamma^{\mathrm{LC}}}$}
\label{sec:exp-Gamma}

In this appendix, we will briefly explain how to derive
$e^{-\Gamma^{\mathrm{LC}}}$ given in eq.(\ref{eq:exp-Gamma}).

Since the interaction vertex $\langle V_{N}|$
given in eq.(\ref{eq:vertex-N}) is defined on the $\rho$-plane
endowed with the flat metric
\begin{equation}
ds^{2} = d\rho d\bar{\rho}~,
\end{equation}
$e^{-\Gamma^{\mathrm{LC}}}$ is the partition function of the
CFT on the $\rho$-plane with this metric.
Here we are not dealing with the superspace formalism, 
and the situation is the same as that in the bosonic string case. 
As explained in Ref.~\cite{Mandelstam:1985ww},
its dependence on $\alpha_{r}$ and the moduli parameters
$\mathcal{T}_{I}$ can be determined through the CFT technique
by evaluating the Liouville action associated with
the conformal mapping (\ref{eq:Mandelstam})
between the $\rho$-plane and the $z$-plane with small circles
around the $N-2$ interaction points $z_{I}$,
the $N$ punctures $Z_{r}$ and $\infty$ excised.
Collecting the contributions from these holes on the $z$-plane,
we obtain
\begin{equation}
-\Gamma^{\mathrm{LC}}
 = \frac{c}{12} \sum_{r=1}^{N}
    \ln \left| \frac{1}{\alpha_{r}}
               \frac{\partial w_{r}}{\partial z} (Z_{r})
         \right|
  -\frac{c}{48} \sum_{I}
     \ln \left( \partial^{2} \rho (z_{I})
                     \bar{\partial}^{2} \bar{\rho} (\bar{z}_{I})
                     \right)
  + \frac{c}{6} \ln \left| \lim_{z\rightarrow \infty}
                            z^{2} \partial \rho (z)
                    \right|~,
\label{eq:Liouville}
\end{equation}
up to an additive constant.
Here $w_{r}$ is defined in eq.(\ref{eq:rho-wr}).
Combined with the relation
\begin{equation}
\ln \left(\frac{\partial w_{r}}{\partial z} (Z_{r}) \right)
= - \bar{N}^{rr}_{00}~,
\end{equation}
the above equation leads to eq.(\ref{eq:exp-Gamma}), 
up to an overall constant, 
which will be discussed in the following.

\subsubsection*{Factorization of $e^{-\Gamma^{\mathrm{LC} }}$}

\begin{figure}[htbp]
\begin{center}
	\includegraphics[width=20.5em]{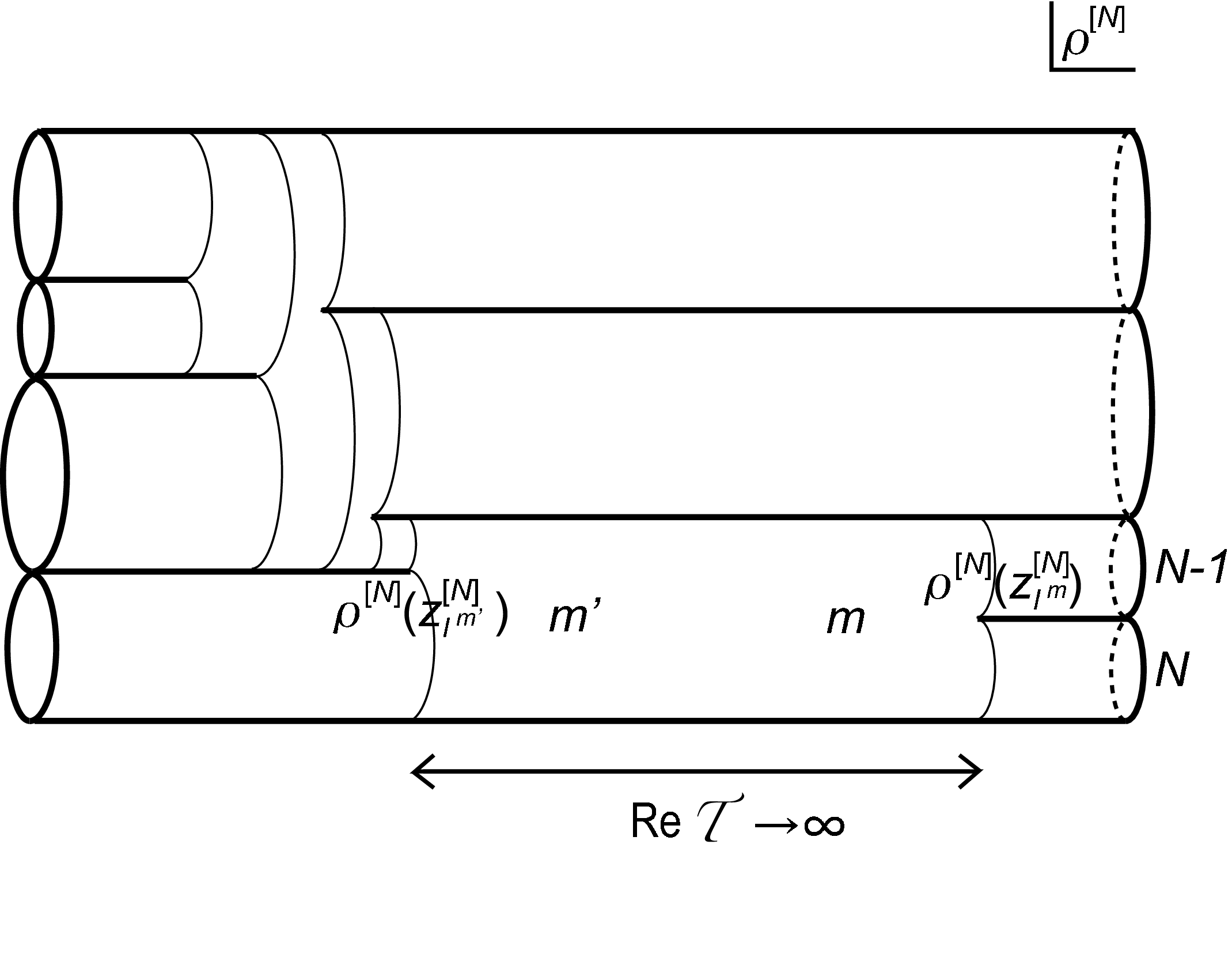}
	\caption{The interaction points for strings $m$ and $m'$
                 are denoted by $\rho^{[N]}(z^{[N]}_{I^{m}})$
                 and $\rho^{[N]} (z^{[N]}_{I^{m'}})$ respectively.
                We examine the limit in which
                 $\mathcal{T}=\rho^{[N]}(z^{[N]}_{I^{m}})
                  -\rho^{[N]} (z^{[N]}_{I^{m'}})
                  \rightarrow \infty$.}
	\label{fig:factorization}
\end{center}
\end{figure}

We would like to show that $e^{-\Gamma^{\mathrm{LC} }}$ given 
in eq.(\ref{eq:exp-Gamma}) has the correct normalization. 
In the $N=3$ case, 
using the relations in appendix \ref{sec:Mandelstam},
one can readily find that
$e^{-\Gamma^{\mathrm{LC}}}$ in eq.(\ref{eq:exp-Gamma}) coincides with
$e^{-\Gamma^{\mathrm{LC}[3]}}$
contained in the three-string vertex (\ref{eq:exp-Gamma-3}).
Now that the $N=3$ case is proved,
the $N \geq 4$ cases are proved
if the factorization
\begin{eqnarray}
e^{-\Gamma^{\mathrm{LC}[N] }}(1,2,\dots,N)
&\stackrel{\mathcal{T}\rightarrow \infty}{\longrightarrow}&
 e^{-\Gamma^{\mathrm{LC}[N-1] }}(1,2,\dots,N-2,m')
\nonumber\\
&&\quad \times
\left( - e^{\frac{c}{12} \frac{ \mathrm{Re} \mathcal{T}}
{|\alpha_{m}| }}\right)
e^{-\Gamma^{\mathrm{LC}[3] }}(m,N-1,N)
\label{eq:factorization}
\end{eqnarray}
is shown for $N\geq 4$. Here
$\mathcal{T}$ is the moduli parameter of the $N$-string
tree diagram depicted in Fig.~\ref{fig:factorization},
and
$e^{-\Gamma^{\mathrm{LC}[n]}}$ denote
the factor (\ref{eq:exp-Gamma}) for the $n$-string case.
$m$ and $m'$ stand for the intermediate string
described in Fig.~\ref{fig:factorization}
and therefore the string-length parameter $\alpha_{m}$ satisfies 
$\alpha_{m} =-\alpha_{N-1} - \alpha_{N}$.
The factor 
$\left(
-e^{\frac{c}{12} \frac{ \mathrm{Re} \mathcal{T}}
                      {|\alpha_{m}| }} \right)$
is the contribution from the propagator
(\ref{eq:propagator}) of
the lowest level state of the intermediate string.

Let us briefly explain how to show eq.(\ref{eq:factorization}).
For this computation,
it is useful to take $Z_{N}=\infty$.
In this gauge, the Mandelstam mapping for the $N$-string case
becomes
\begin{equation}
\rho^{[N]}(z) = \sum_{r=1}^{N-1} \alpha_r \ln (z-Z_r)~.
\label{eq:Mandelstam-infinity}
\end{equation}
We also introduce
$\rho^{[N-1]}(z) = \sum_{r=1}^{N-2} \alpha_r \ln (z-Z_r)$.
In terms of $\rho^{[N]}(z)$,
$e^{-\Gamma^{\mathrm{LC}[N] }}$ is described as
\begin{equation}
e^{-\Gamma^{\mathrm{LC}[N] }}(1,2,\dots,N)
=\mathrm{sgn}\left( \prod_{r=1}^N \alpha_r\right)
\left|\frac{\alpha_N}{
\prod_{r=1}^{N-1} \alpha_r }
\right|^{\frac{c}{12}}
\prod_{I}\left|\partial^{2} \rho^{[N]}(z_I^{[N]}) 
\right|^{- \frac{c}{24}}
e^{-\frac{c}{12} \sum_{r=1}^N \mathrm{Re}\bar{N}_{00}^{rr[N]}   }~,
\label{eq:exp-Gamma-N2}
\end{equation}
where
\begin{equation}
\mathrm{Re} \, \bar{N}_{00}^{rr[N]}  
= \frac{\tau_0^{(r)[N]} }{\alpha_r}
 - \sum_{s\neq r, N}\frac{\alpha_s}{\alpha_r}
 \ln |Z_r- Z_s|~,
 \qquad 
\mathrm{Re} \, \bar{N}_{\; 00}^{NN[N]}  
= \frac{\tau_0^{(N)[N]} }{\alpha_N}
\end{equation}
for $r=1,2,\dots ,N-1$.

Let us take the limit $Z_{N-1} \to \infty$ 
with the other $Z_r$'s  kept fixed.
In this limit, the moduli parameter
$\mathcal{T}$ becomes infinity,
while the other ones 
tend to the moduli parameters determined
by the Mandelstam mapping $\rho^{[N-1]}(z)$.
Taking the limit $Z_{N-1}\to \infty$ in eq.(\ref{eq:exp-Gamma-N2}),
one can find that the factorization (\ref{eq:factorization})
indeed holds.


\newpage


\begin{thebibliography}{99}




\bibitem{Mandelstam:1974hk}
  S.~Mandelstam,
  ``Interacting String Picture of the Neveu-Schwarz-Ramond Model,''
  Nucl.\ Phys.\  B {\bf 69}, 77 (1974).


\bibitem{Sin:1988yf}
  S.~J.~Sin,
  ``GEOMETRY OF SUPER LIGHT CONE DIAGRAMS AND LORENTZ INVARIANCE
   OF LIGHT CONE STRING FIELD THEORY. 2. CLOSED NEVEU-SCHWARZ STRING,''
  Nucl.\ Phys.\  B {\bf 313}, 165 (1989).


\bibitem{Mandelstam:1985wh}
  S.~Mandelstam,
  ``Interacting String Picture Of The Fermionic String,''
  Prog.\ Theor.\ Phys.\ Suppl.\  {\bf 86}, 163 (1986).




\bibitem{Greensite:1986gv}
  J.~Greensite and F.~R.~Klinkhamer,
  ``New Interactions For Superstrings,''
  Nucl.\ Phys.\  B {\bf 281}, 269 (1987);
  ``CONTACT INTERACTIONS IN CLOSED SUPERSTRING FIELD THEORY,''
  Nucl.\ Phys.\  B {\bf 291}, 557 (1987);
  ``Superstring Amplitudes And Contact Interactions,''
  Nucl.\ Phys.\  B {\bf 304}, 108 (1988).


\bibitem{Green:1987qu}
  M.~B.~Green and N.~Seiberg,
  ``CONTACT INTERACTIONS IN SUPERSTRING THEORY,''
  Nucl.\ Phys.\  B {\bf 299}, 559 (1988).







\bibitem{Wendt:1987zh}
  C.~Wendt,
  ``SCATTERING AMPLITUDES AND CONTACT INTERACTIONS IN WITTEN'S
    SUPERSTRING FIELD THEORY,''
  Nucl.\ Phys.\  B {\bf 314}, 209 (1989).


\bibitem{Berkovits:1985ji}
  N.~Berkovits,
  ``Calculation Of Scattering Amplitudes For The Neveu-Schwarz
    Model Using Supersheet Functional Integration,''
  Nucl.\ Phys.\  B {\bf 276} (1986) 650;
  ``Supersheet Functional Integration And The Interacting
    Neveu-Schwarz String,''
  Nucl.\ Phys.\  B {\bf 304}, 537 (1988);
  ``SUPERSHEET FUNCTIONAL INTEGRATION AND THE CALCULATION
    OF NSR SCATTERING AMPLITUDES INVOLVING ARBITRARILY MANY
    EXTERNAL RAMOND STRINGS,''
  Nucl.\ Phys.\  B {\bf 331}, 659 (1990).


\bibitem{Aoki:1990yn}
  K.~Aoki, E.~D'Hoker and D.~H.~Phong,
  ``UNITARITY OF CLOSED SUPERSTRING PERTURBATION THEORY,''
  Nucl.\ Phys.\  B {\bf 342}, 149 (1990).




\bibitem{Mandelstam:1973jk}
  S.~Mandelstam,
  ``Interacting String Picture of Dual Resonance Models,''
  Nucl.\ Phys.\  B {\bf 64}, 205 (1973).




\bibitem{LeClair:1988sp}
  A.~LeClair, M.~E.~Peskin and C.~R.~Preitschopf,
  ``String Field Theory on the Conformal Plane. 1. 
    Kinematical Principles,''
  Nucl.\ Phys.\  B {\bf 317}, 411 (1989).


\bibitem{Mandelstam:1985ww}
  S.~Mandelstam,
  ``The Interacting String Picture And Functional Integration,''
  in {\it Workshop on Unified String Theories,\/}
  eds.~ M.B.~Green and D.J.~Gross
  (World Scientific, Singapore, 1986), p46.





\bibitem{Friedan:1985ge}
  D.~Friedan, E.~J.~Martinec and S.~H.~Shenker,
  ``Conformal Invariance, Supersymmetry And String Theory,''
  Nucl.\ Phys.\  B {\bf 271}, 93 (1986).

\end{thebibliography}
\end{document}